# Inconsistency of the non-standard definition of work

Jose M. G. Vilar[1] and J. Miguel Rubi[2]

[1]Computational Biology Program, Memorial Sloan-Kettering Cancer Center, 1275 York Avenue, New York, NY 10021
[2]Departament de Fisica Fonamental, Universitat de Barcelona, Diagonal 647, 08028 Barcelona, Spain

**Abstract**
We show that the recently postulated non-standard definition of work proportional to force variation rather than to displacement [A. Imparato and L. Peliti, cond-mat arXiv:0706.1134v1] is thermodynamically inconsistent at both microscopic and macroscopic scales and leads to non-physical results, including free energy changes that depend on arbitrary parameters.

In a recent preprint, Imparato and Peliti [1] contend that the standard textbook definition of work,

$$\text{Work} = \text{Force} \times \text{Displacement}, \qquad (1)$$

is the incorrect expression for the work performed by a force on a system. Their criticism of the standard definition arises as a response to our recent results [2] that show that it is not possible to use changes in the Hamiltonian to compute thermodynamically consistent free energy differences when work is not connected to changes in the Hamiltonian. They argue in the opposite direction, namely, that free energy differences should be computed with a new definition of work that restores the work-Hamiltonian connection. In their definition of work, denoted here as Work$_{IP}$ or $W_{IP}$, the force $f$ and position $x$ have their roles exchanged, e.g. $dW_{IP} = -x df$ (Eq. 16 of reference [1]). Here we show that this non-standard definition of work is thermodynamically inconsistent at both microscopic and macroscopic scales and leads to non-physical results, including free energy changes that depend on arbitrary parameters.

To set a common ground, let us consider a system in a thermostat at a temperature $T$ described by the Hamiltonian $H_0(x) = \tfrac{1}{2}kx^2$, where $k$ is a constant and $x$ is the coordinate, in the presence of an external time-dependent force $f \equiv f(t)$. The Hamiltonian of the system plus force is

$$H(x;t) = \tfrac{1}{2}kx^2 - fx. \qquad (2)$$

This Hamiltonian describes many types of systems around a minimum of potential in the presence of a homogeneous force that does not depend on the coordinate $x$. It has been used to analyze, mechanical, magnetic, electrical, and chemical systems at both microscopic and macroscopic scales, accounting for situations as diverse as a microscopic molecule that is stretched in the linear regime all the way up to a macroscopic spring that follows hook's law. At equilibrium, the effects of thermal fluctuations can be obtained directly through the canonical distribution $P(x;t) = e^{-H(x;t)/k_BT}/Z$, where $Z = \int e^{-H(x;t)/k_BT}dx$ is the partition function.

The macroscopic spring is an interesting example because it allows one to compute work and free energy changes by using just elementary physics (cf. Section 7-7 of Ref. [3]) and provides a benchmark for the limit of negligible thermal fluctuations. In this case, the reversible, quasi-static work performed by the external force is $dW_{rev} = kxdx$ and the energy stored in the spring is

$$\Delta G = \int_0^{x_{eq}=f/k} dW_{rev} = \frac{1}{2}kx_{eq}^2 = \frac{f^2}{2k}, \qquad (3)$$

which coincides with the free energy change. If the force is instantaneously switched on at time $t = 0$ (e.g. $f = f_0\Theta(t)$, with $\Theta(t)$ the Heaviside step function), the work performed $dW = fdx$ leads to

$$W = f\int_0^{x_{t\to\infty}=f/k} dx = fx_{eq} = \frac{f^2}{k}. \qquad (4)$$

Imparato and Peliti postulate that for the above Hamiltonian [Eq. (2)] the definition of work that should be used is $dW_{IP} = -xdf$ (Eq. 16 of Ref. [1]). The first question to address concerns the basic properties of Work$_{IP}$. For instance, if the external force is instantaneously switched on at time zero, one has

$$W_{IP}\Big|_{t=-\varepsilon/2}^{t=\varepsilon/2} = \int_{-\varepsilon/2}^{\varepsilon/2} -x\frac{df}{dt}dt = -x_0 f_0, \qquad (5)$$

where $x_0$ is the initial value of the coordinate, which would imply that a finite force could perform a constant finite amount of work during an arbitrarily small time interval $\varepsilon$. Work$_{IP}$ also leads to the result that the force would stop performing any work beyond $t = 0^+$, even though the force remains present and acting on the evolving system. In the case of the macroscopic spring we have discussed, the force does not perform any Work$_{IP}$ at all because $x_0 = 0$. However, Work$_{IP}$ is not translationally invariant and it can take any arbitrary value just by changing the reference system. Most remarkably, Work$_{IP}$ can be different from zero for a system that does not change its state in response to the external applied force.

The properties of Work$_{IP}$, as we have seen, strongly contrast with those of work. *"In the scientific sense no work is ever done unless the force succeeds in producing motion in the body on which it acts. A pillar supporting a building does no work; a man tugging at a stone, but failing to move it, does no work. In the popular sense we sometimes say that we are doing work when we are simply holding a weight or doing anything else which results in fatigue; but in physics the word "work" is used to describe* *not the effort put forth but the effect accomplished*" (quoted from Chapter VII of Ref. [4]). In this regard, irrespective of whether the system is macroscopic or microscopic, different types of work are related to different types of motion: mechanical work to motion of bodies, electric work to motion of charges, paramagnetic work to orientation of magnetic dipoles, and chemical work to rearrangement of atoms between different chemical species.

From a thermodynamic point of view, work is a form of energy transfer connected to other forms of energy transfer, including heat. As such, it is a measurable quantity that has a precise determined value. To give support to their course of reasoning, Imparato and Peliti quoted a paragraph from Tolman's classic book on 'The Principles of Statistical Mechanics' [5], but the paragraph as quoted in Ref. [1] is missing not only the context of the chapter where it belongs to but also an important clarifying sentence. The omitted sentence, which was replaced by "[…]" in Ref. [1], reads in Tolman's book as follows: "*As a simple typical example, when the volume of a system is varied by the amount δv, we set the work done and the energy thereby transferred equal to pδv, where the generalized force p is the pressure exerted by the system.*" Tolman's book, in fact, uses the standard definition of work, generalized force ($p$) times generalized displacement ($\delta v$).

What are the consequences of using Work$_{IP}$ instead of work for free energy calculations? It is obtained in Ref. [1] from Work$_{IP}$ that the free energy change of stretching a spring-like system described by the Hamiltonian of Eq. (2) at a temperature $T$ is negative: $\Delta G_{IP} = -\frac{f^2}{2k}$ (Eq. (21) of Ref. [1]). This result is at odds with the thermodynamic hallmark that non-spontaneous processes lead to positive free energy changes and with previous studies, which obtained positive values for stretching macromolecules [6]. Note also that the resulting $\Delta G_{IP}$ does not depend on the temperature and should then hold for any system described by the Hamiltonian of Eq. (2), including a macroscopic spring, for which $\Delta G$ is positive [Eq. (3)]. The use of Work$_{IP}$ then raises questions not only about its microscopic suitability but also about its macroscopic limit, which does not recover the elementary physics textbook result.

The negative value of $\Delta G_{IP}$ is argued in Ref. [1] to be the result of including the free energy of the potential associated with the external force. A question that would remain then is how to calculate the free energy of the system, which is positive, from that of the system plus the external force used to measure it, which leads to negative values of $\Delta G_{IP}$. In general, if one is interested in characterizing the properties of a system, it is not convenient to have the particular properties of the applied external force embedded into the results.

A thorough analysis, however, reveals that the approach followed to obtain Work$_{IP}$ provides undetermined results. Consider for instance a more general Hamiltonian

$$H(x) = \tfrac{1}{2}kx^2 - f(x - \gamma), \tag{6}$$

where $\gamma$ is a constant parameter that does not affect in any way the dynamics of the system. The line of reasoning followed to postulate Work$_{IP}$ is that it is the change in the Hamiltonian upon variation of the force: $dW_{IP} = (\partial H / \partial f) df$. In this case, one has $\partial H / \partial f = -(x - \gamma)$, which leads to

$$dW_{IP} = -(x - \gamma) df . \tag{7}$$

Therefore, Work$_{IP}$ depends in general on a non-physical parameter $\gamma$, which cannot be measured and which does not affect the dynamics of the system. Consequently, Work$_{IP}$ is determined not by the actual physical system but by the particular mathematical approach used to describe it.

Imparato and Peliti define the reversible work from the average $dW_{revIP} = \langle \partial H / \partial f \rangle_f df$, which leads them to $\Delta G_{IP} = \int_0^f dW_{revIP}$. Integration of the averaged reversible Work$_{IP}$ over the force results in

$$\Delta G_{IP} = -\int df (\langle x \rangle_f - \gamma) = -\frac{f^2}{2k} + \gamma f , \tag{8}$$

which shows that free energy differences computed from Work$_{IP}$ depend also on the undetermined parameter $\gamma$. Thermodynamic free energy differences between two states, however, should be completely specified as their sign determine whether or not processes happen spontaneously.

It should be emphasized that the definition of reversible work assumed in Ref. [1], $W_{revIP} = \int_0^f \langle \partial H / \partial f \rangle_{f'} df'$, cannot be applied in general over a single microscopic trajectory because it is an ensemble average. By definition the macroscopic reversible work is not a fluctuating quantity, but the work over a microscopic quasi-static, reversible trajectory does indeed fluctuate. Averaging over the work $W_{qs}$ done by the external force over a quasi-static microscopic trajectory leads to

$$\Delta G = \langle W_{qs} \rangle \neq -k_B T \ln \langle e^{-W_{qs}/k_B T} \rangle , \tag{9}$$

irrespective of whether the reversible work is considered microscopically as $W_{rev} = W_{qs}$ or macroscopically as $W_{rev} = \langle W_{qs} \rangle$. This elementary inequality shows that the Jarzynski equality cannot be used in general to compute free energy changes from work exponential averages. It has been shown that Work$_{IP}$ follows exactly the Jarzynski equality for systems with Langevin dynamics [7]. Therefore, the properties of Work$_{IP}$ are such that for quasi-static, reversible trajectories one has

$$W_{revIP} = \int_0^f \langle \partial H / \partial f \rangle_{f'} df' = \Delta G_{IP} = -k_B T \ln \left\langle \exp\left(-\frac{1}{k_B T} \int_0^f (\partial H / \partial f) df'\right) \right\rangle . \tag{10}$$

We have recently shown [2] that it is not possible to obtain thermodynamically consistent free energy differences between two states from changes in the Hamiltonian when changes in the Hamiltonian do not represent the work done on the system. Redefinition of work, as proposed in Ref. [1], can keep the mathematical self-compatibility of expressions that rely on the work-Hamiltonian connection but does not solve their physical inconsistencies. In particular, both the non-standard definition of work and free energy changes calculated from it are not univocally determined since they depend on arbitrary mathematical parameters that do not affect the dynamics of the system. These inconsistencies are not present when the textbook definition of work, *Work = Force × Displacement*, is used.